\documentclass[aps,prl,groupedaddress,twocolumn,showpacs,preprintnumbers,amsmath,amssymb,floatfix,superscriptaddress]{revtex4}

\pdfoutput=1

\usepackage{graphicx,color}
\usepackage{dcolumn}
\usepackage{bm}
\usepackage{color}
\usepackage{amsmath}
\usepackage{multirow}

\newcommand{\E}{EuTiO$_3$}
\newcommand{\Sr}{SrTiO$_3$}

\begin{document}

\title{\boldmath Intrinsic Structural Disorder and the Magnetic Ground State in Bulk EuTiO$_3$}

\author{A.P. Petrovi\'c}
\affiliation{Division of Physics and Applied Physics, Nanyang Technological University, 637371 Singapore}
\author{Y. Kato}
\affiliation{Theoretical Division, Los Alamos National Laboratory, CNLS and T-4, Los Alamos, New Mexico 87545, USA}
\author{S.S. Sunku}
\affiliation{Division of Physics and Applied Physics, Nanyang Technological University, 637371 Singapore}
\author{T. Ito}
\affiliation{Correlated Electron Engineering Group, AIST, Ibaraki 305-8562, Japan}
\author{P. Sengupta} 
\affiliation{Division of Physics and Applied Physics, Nanyang Technological University, 637371 Singapore}
\author{L. Spalek}
\affiliation{Cavendish Laboratory, University of Cambridge, Cambridge CB3 0HE, United Kingdom}
\affiliation{Department of Physics, University of Crete and FORTH, 71003 Heraklion, Greece}
\author{M. Shimuta}
\affiliation{Department of Physics, Waseda University, Tokyo 169-8555, Japan}
\author{T. Katsufuji}
\affiliation{Department of Physics, Waseda University, Tokyo 169-8555, Japan}
\author{C.D. Batista}
\affiliation{Theoretical Division, Los Alamos National Laboratory, CNLS and T-4, Los Alamos, New Mexico 87545, USA}
\author{S. Saxena}
\affiliation{Cavendish Laboratory, University of Cambridge, Cambridge CB3 0HE, United Kingdom}
\author{C. Panagopoulos} 
\affiliation{Division of Physics and Applied Physics, Nanyang Technological University, 637371 Singapore}
\affiliation{Cavendish Laboratory, University of Cambridge, Cambridge CB3 0HE, United Kingdom}
\affiliation{Department of Physics, University of Crete and FORTH, 71003 Heraklion, Greece}

\date{\today}

\begin{abstract}

The magnetic properties of single-crystal EuTiO$_3$ are suggestive of nanoscale disorder below its cubic-tetragonal phase transition.  We demonstrate that electric field cooling acts to restore monocrystallinity, thus confirming that emergent structural disorder is an intrinsic low-temperature property of this material.  Using torque magnetometry, we deduce that tetragonal EuTiO$_3$ enters an easy-axis antiferromagnetic phase at 5.6~K, with a first-order transition to an easy-plane ground state below 3~K.  Our data is reproduced by a 3D anisotropic Heisenberg spin model.  

\end{abstract}


\maketitle

Perovskite titanates $A$TiO$_3$ are among the brightest stars in the rapidly-developing field of oxide electronics.  Their dielectric and transport properties are easily modulated by epitaxial strain, field-effect or chemical doping~\cite{Ohtomo-2002,Chung-2004,Kozuka-2009,Lee-2010}, hence facilitating their integration with conventional charge-based electronics.  However, the past decade has also seen an intensive search for a multiferroic perovskite with strong magnetoelectric coupling, which would enjoy numerous spintronic applications.  The spotlight soon fell on {\E}, whose large magnetic moment ($S$ = 7/2 per unit cell) and quantum paraelectricity suggest that it is a magnetic analogue of {\Sr}, the ``workhorse'' oxide for electronic devices.  Although {\E} has been known as a $G$-type antiferromagnet below T$_N$~$\sim$~5.5~K since the 1960s~\cite{McGuire-1966}, its magnetoelectric properties were only revealed in 2001 by a 7$\%$ drop in the dielectric constant at T$_N$~\cite{Katsufuji-2001}.  Early efforts to model {\E} shared one critical feature: an assumption of cubic crystal symmetry throughout the phase diagram~\cite{Fennie-2006,Ranjan-2007,Shvartsman-2010}.  The recent discovery of a cubic-tetragonal transition at 283~K in powdered samples~\cite{Bussmann-Holder-2011} obliges us to revise our views, since tetragonal symmetry pushes {\E} closer to a ferromagnetic/ferroelectric (FM/FE) phase boundary~\cite{Fennie-2006} and permits the breakage of spatial inversion symmetry.  Theoretical~\cite{Bettis-2012} and X-ray diffraction (XRD) analysis~\cite{Allieta-2012} suggests that tetragonal {\E} may be disordered, thus opening the door to local symmetry-breaking and electric dipole formation via the Dzyaloshinskii-Moriya (DM) interaction.  More generally, nanoscale heterogeneity limits applications for many complex oxides~\cite{Dagotto-2005}, since functional electronic materials must display phase purity over lengthscales greater or equal to their intended device dimensions.  It is therefore an urgent priority to identify methods for controlling or suppressing such disorder.  In this Letter, we employ the magnetic anisotropy of {\E} as a structural probe, hence showing that monocrystalline cubic {\E} undergoes a transition to an intrinsically nano-disordered tetragonal phase with an easy-plane AF ground state.  However, the disorder may be reduced by cooling under an electric field.  This mechanism of homogeneity control may prove crucial for phase management in future devices made from elastically hard oxides.  

Our experiments were performed on {\E} single crystals, grown in a floating-zone furnace~\cite{Katsufuji-1999}.  High-symmetry axes were determined by XRD in the cubic phase at 300~K and the crystals cut into cuboids with faces parallel to (100), (010) and (001).  The cutting/alignment error is $\sim\pm$~5$^{\circ}$, a figure supported by our torque data.  We stress that at 300~K, our samples are perfectly monocrystalline with no evidence for twinning.  Geometric demagnetization corrections~\cite{Aharoni-1998} have been applied to all our data, except the specific heat and torque.  Our results have been verified using four separate crystals, all of which display similar behaviour.

\begin{figure}[htbp]
\centering
\includegraphics [width=8.5cm,clip] {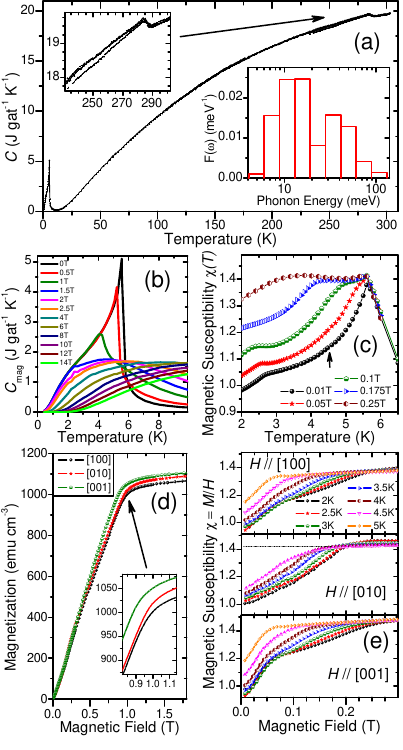}
\caption{\label{Fig1} (a) Zero-field heat capacity $C(T)$ in {\E}.  Top left: zoom on the structural transition at 283~K, showing two divergent traces generated by consecutive thermal relaxations.  Lower right: PDOS extracted from the specific heat (see ref.~\cite{Petrovic-2010} for details).  (b) Low temperature magnon heat capacity $C_{\mathrm{mag}}$, illustrating the field-suppression of AF order and spin gap formation.  The absence of the spin-flop and spin realignment transitions from $C(T,H)$ is unsurprising, since the entropy change is extremely small and short-pulse relaxation calorimetry is notoriously insensitive to first-order transitions~\cite{Lashley-2003}.  (c) dc magnetic susceptibility $\chi(T)$ for a range of applied fields, revealing the transition to $ab$-plane order below $\sim$~3~K, a ``shoulder'' due to the spin-flop and a peak at T$_N$.  A faint anomaly is also visible in low fields around 4.5~K (black arrow).  (d) dc magnetization $M(H)$ along the crystal axes.  Small anisotropies in $m_{sat}$ arise from an anisotropic $g$-tensor in a tetragonal system; the fact that our observed anisotropy is quasi-3D rather than 2D is indicative of disorder.  (e) Spin-flop transitions in $\chi(H)$ for fields parallel to each axis.  In the absence of a spin-flop, $\frac{\delta\chi}{{\delta}H}$ should remain constant down to $H=0$ (as illustrated by a dashed line for ${\bm H}$~//~[010]).}  
\end{figure}

The structural transition in {\E} is visible as a kink in the heat capacity $C(T)$ at $T_c$~=~283~K (Fig.~\ref{Fig1}a).  As we sweep the temperature below $T_c$, repeated thermal relaxations yield two divergent traces in $C(T)$, the lower of which is generated by the first relaxation at a given temperature and vanishes below $\sim$~240~K.  This trend is reproducible for all our crystals, but absent from reference materials tested on the same calorimeter and hence not an experimental artifact.  Such behaviour is suggestive of phase metastability in this temperature range.  Interestingly, XRD indicates a lower $T_c$ of $\sim$~235~K~\cite{Allieta-2012}, coinciding with the disappearance of our lower trace.  We extract the temperature-integrated phonon density of states (PDOS) from $C(T)$ and note the broad spectral weight distribution below 20~meV: this corresponds to the TO1 phonon softening from 120~cm$^{-1}$ to 80~cm$^{-1}$~\cite{Goian-2009}.  Subtracting a $\beta_1T^3 + \beta_2T^5$ phonon background from $C(T)$ at low temperature, we obtain the magnon heat capacity $C_{\mathrm{mag}}$ (Fig.~\ref{Fig1}b).  A large jump at $T_N$~=~5.6~K indicates a second-order AF transition, similar to existing powder~\cite{Bussmann-Holder-2011} and nanoparticle~\cite{Wei-2011} data.  Integrating $C_{\mathrm{mag}}/T$ from zero to 5.6~K yields a total entropy of 69.4~$\pm$~1~mJgat$^{-1}$K$^{-1}$ which agrees with the theoretical value $Nk_B$ln$(2S+1)$ = 69.8~mJgat$^{-1}$K$^{-1}$ and discourages the possibility of any further structural transition at T~$\leq$~5.6~K.  Magnetic fields suppress $T_N$, inducing a $T=0$ transition (at $H=H_{sat} \simeq 1$T) to saturated paramagnetism, with an energy gap for spin excitations growing linearly in $H-H_{sat}$.  This confirms the phase purity of our samples, since any oxygen deficiency should drive a zero-field crossover from AF to FM order~\cite{Kugimiya-2007}.  

Further complexity emerges in the dc magnetic susceptibility $\chi=M/H$ (Fig.~\ref{Fig1}c), where three energy scales may be identified.  The peak in $\chi(T)$ at $T_N$ is robust at low fields; however a ``shoulder'' forms below $T_N$, broadening and moving to lower temperature as $H$ increases.  A third anomaly is visible below 3~K, vanishing above $H~\sim$~0.1~T.  The magnetization saturates at a similar field $H_{sat}~\sim$~1~T along all three axes (Fig.~\ref{Fig1}d), with a moment $m_{sat}$~=~7.08~$\pm$~0.2~$\mu_B$/unit cell.  At low fields, small anomalies in $M(H)$ are visible and $\chi(H)$ falls as $H$ tends to zero (Fig.~\ref{Fig1}e).  The temperature and field-dependence of these non-linearities in $M(H)$ correlate with the ``shoulder'' in $\chi(T)$, leading us to identify them as spin-flop transitions.  However, a single-crystal AF with tetragonal symmetry should only exhibit a spin-flop when ${\bm H}$ is parallel to the ``easy'' axis (and no spin-flop should occur in an easy-plane system).  In contrast, our crystals consistently display spin-flops at similar fields ($H_{sf}$~=~0.22~$\pm$~0.04~T at 2~K) along all three axes .  

\begin{figure}[t]
\centering
\includegraphics [width=8.5cm,clip] {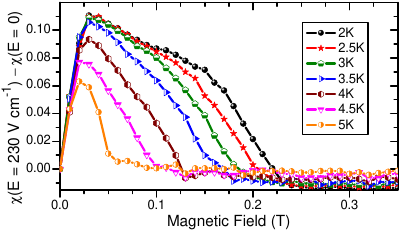}
\caption{\label{Fig2} Electric field-induced spin-flop suppression in an {\E} crystal cooled from $T$~=~80~K under $E$~=~230~V~cm$^{-1}$ (${\bm E}{\perp}{\bm H}$).  To apply the field, gold contacts were sputtered on opposing crystal faces and wires attached with silver epoxy.}  
\end{figure}

To resolve this conflict, let us suppose that {\E} is indeed disordered below $T_c$, as suggested by our $C(T)$ and $M(H)$ data as well as other recent work~\cite{Allieta-2012,Bettis-2012}.  In this case, tetragonal {\E} may be considered as effectively polycrystalline.  Our experiments do not permit us to estimate the lengthscale of this disorder; however, we note that spontaneous twinning is observed in elastically soft {\Sr} with domain size below 1~$\mu$m~\cite{Chrosch-1998}.  The presence of a spin-flop in a 200~nm {\E} film also suggests that true nanoscale disorder exists in this material~\cite{Lee-2009}.  Nevertheless, it should be possible to reduce this disorder by cooling {\E} with an electric field ${\bm E}$ applied parallel to a cubic crystal axis: this prevents softening for phonon wavevectors \textbf{k}//${\bm E}$~\cite{Worlock-1967} and hence ensures that the tetragonal $c$-axis is perpendicular to ${\bm E}$.  It is crucial to $E$-field-cool (EFC) from above 50~K, since below this temperature the phonon softening terminates, quantum fluctuations dominate the low-energy phonon spectrum and the lattice structure is presumably locked.  Fig.~\ref{Fig2} shows the effect of EFC on $\chi(H)$ for a {\E} crystal cooled from 80~K: the spin-flop is clearly suppressed, implying that an increased proportion of our crystal now exhibits $c$//${\bm H}$.  This confirms the presence of structural disorder and its influence on the magnetic ground state.   

\begin{figure}[ht!]
\centering
\includegraphics [width=8.5cm,clip] {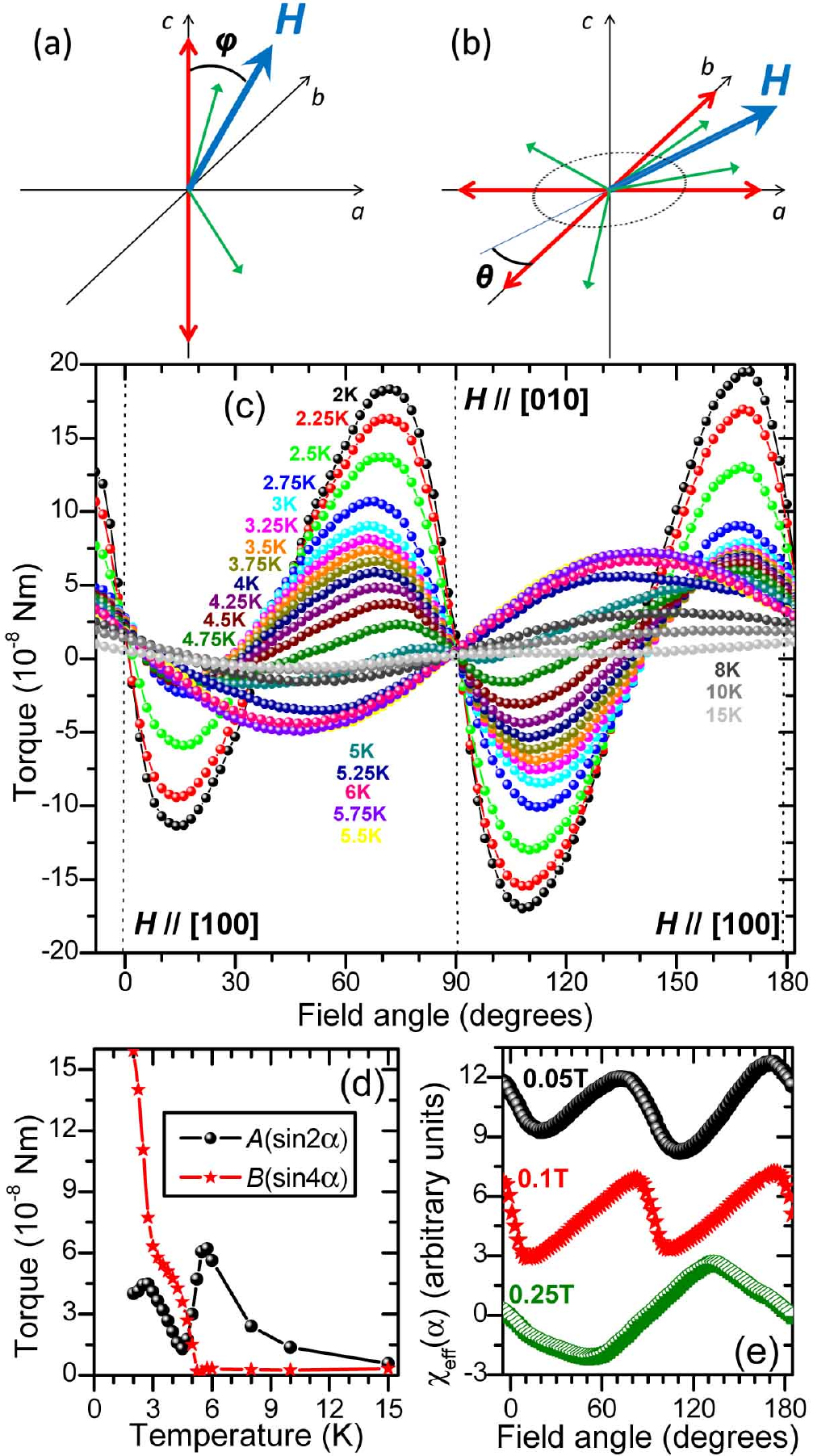}
\caption{\label{Fig3} (a,b) Torque experimental schematic, with easy axes shown in red, AF sublattice moments in green and field ${\bm H}$ in blue.  (c) Temperature dependence of $\tau(\alpha)$~//~[001] for  $H$~= 0.05~T.  Rotating ${\bm H}$ in the (100) and (010) planes gives similar results.  (d) Temperature dependence of the separated sin2$\alpha$ and sin4$\alpha$ contributions to the total torque shown in (c).  (e) Field-evolution of the effective susceptibility $\chi(\alpha)\equiv\tau(\alpha)/H^2$ revealing the spin-flop transition.}  
\end{figure}

Since europium compounds exhibit significant neutron absorption and are hence poorly-suited to neutron diffraction, torque magnetometry is our best alternative to elucidate the spin structure~\cite{Meijer-1998,Herak-2010}.  An easy-axis two-sublattice AF (Fig.~\ref{Fig3}a) will generate an out-of-plane torque signal $\tau(\phi)~\propto~\mathrm{sin}2\phi$ when a field ${\bm H}$ rotates in a plane containing the easy axis; however $\tau~=~0$ if ${\bm H}$ lies in the plane perpendicular to the easy axis.  Conversely, an easy-plane AF (Fig.~\ref{Fig3}b) will display $\tau(\theta)~\propto~\mathrm{sin}4\theta$ when ${\bm H}$ is rotated in the easy plane and $\tau(\phi)~\propto~\mathrm{sin}2\phi$ when ${\bm H}$ is rotated through any plane containing the hard axis ($\phi$ and $\theta$ are the polar and azimuthal angles respectively).  Anisotropic paramagnetism above $T_N$ will also generate a $\tau(\phi)~\propto~\mathrm{sin}2\phi$ signal.  

Our magnetization data imply that {\E} crystals contain AF domains with the easy-axis pointing along all three crystallographic axes. Therefore,  the torque should behave as $\tau(\alpha)~=~A$sin$2\alpha~+~B$sin$4\alpha$, where $\alpha$ is the field angle in a plane defined by any pair of crystal axes.  (The demagnetization factor also contributes to $\tau(\alpha)$ with 180$^\circ$ periodicity, but it is temperature invariant and does not affect our conclusions.)  A series of $\tau(\alpha)$ curves for 2~K~$\geq~T~\geq$~15~K is shown in Fig.~\ref{Fig3}c, with $H$~=~0.05~T rotating in the (001) plane: the transition from sin2$\alpha$ behaviour above $T_N$ to sin4$\alpha$ at low temperature is clearly visible.  Fitting our $\tau(\alpha)$ curves to obtain the sin2$\alpha$ and sin4$\alpha$ weight parameters $A$ and $B$, we note that $B$ only appears below $T_N$, rising rapidly below 3~K with a corresponding drop in $A$.  The only mechanism which can generate this behaviour is a change from easy-axis to easy-plane symmetry: we deduce that a first-order phase transition from $c$-axis to $ab$-plane AF ordering occurs close to 2.75~K.  Upon increasing $H$ (Fig.~\ref{Fig3}e), $\tau(\alpha)$ changes shape from sinusoidal to ``sawtooth'', indicating a spin-flop~\cite{Herak-2010}.  At $H$~=~0.25~T, the approximate sin2$\alpha$ dependence is characteristic of a spin-flopped AF.  

\begin{figure*}[ht!]
\centering
\includegraphics*[width=16cm,clip] {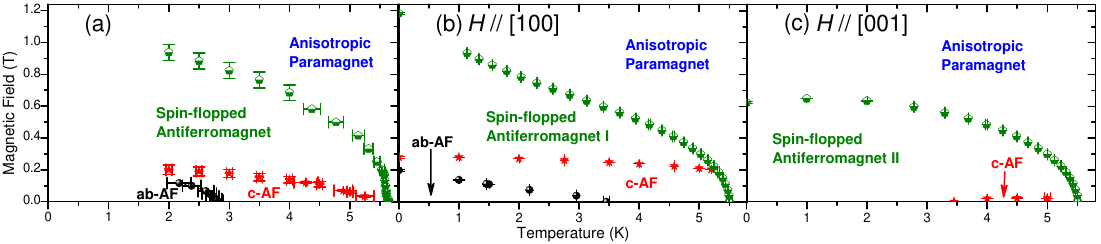}
\caption{\label{Fig4} (a) Experimentally-determined magnetic phase diagram of {\E}.  (b,c) Two-sublattice mean-field phase diagrams for ${\bm H}$~//~[100] and [001] (see text for parameters).  $ab$-AF refers to zero-field AF order along the [110] or [1${\bar 1}$0] axes, while $c$-AF describes AF order along [001].  For the spin-flopped AF I phase, the spins order in the $ab$ plane with the staggered component of the magnetization perpendicular to the applied field.  In contrast, spin-flopped AF II corresponds to AF ordering along [110] or [1${\bar 1}$0], with a uniform magnetization component along the $c$-axis.  The intrinsic structural disorder implies that our experimentally-determined diagram (a) should be a superposition of (b) and (c).  Experimental evidence for the small pocket of $c$-AF in (c) is provided by the faint anomaly visible at low fields in $\chi(T)$ (indicated by an arrow in Fig.~\ref{Fig1}c.)} 
\end{figure*}

We summarise our low-temperature data in Fig.~\ref{Fig4}a.  For zero-EFC {\E}, little or no anisotropy is observed for different magnetic field orientations, as expected for an effective polycrystal.  This phase diagram indicates that single-ion anisotropy terms induced by the tetragonal crystal field play an important role at low temperature.  To model the effect of these terms, we consider a classical Heisenberg model with nearest-neighbour ($J_1$) and next-nearest-neighbour ($J_2$) exchange interactions
\begin{eqnarray}
\mathcal{H} &=& 
J_1 \sum_{\langle {\bm r}, {\bm r}'\rangle} {\bm S}_{\bm r} \cdot {\bm S}_{\bm r'} +
J_2 \sum_{\langle\langle {\bm r}, {\bm r}' \rangle\rangle} {\bm S}_{\bm r} \cdot {\bm S}_{\bm r'}
\nonumber\\
&&+\sum_{\bm r} V({\bm s}_{\bm r})
- \sum_{\bm r} {\bm B} \cdot {\bm S}_{\bm r},
\end{eqnarray}
together with a single-ion anisotropy potential 
\begin{eqnarray}
V({\bm s})=-C_2 (S_x^2+S_y^2)-C_4 (S_x^4+S_y^4),
\end{eqnarray}
where $S_x={\bm S}\cdot(1,1,0)/\sqrt{2}$, and $S_y={\bm S}\cdot(-1,1,0)/\sqrt{2}$.
${\bm S}_{\bm r}$ represents classical spin at ${\bm r}$ with $|{\bm S}_{\bm r}|=7/2$.
The exchange parameters were determined by fitting the measured $T=0$ saturation field $ H_{sat}= 12 S J_1 / g \mu_B  \sim$~1~T and Curie-Weiss temperature $T_{CW} = -2S^2 (J_1 + 2 J_2)\simeq$ 4~K \cite{McGuire-1966} ($g~\simeq~2$ is the gyromagnetic factor), resulting in $J_1S^2=0.31$~K and $J_2S^2=-1.22$~K.  This yields a mean field $T_N = 2S^2(J_1-2J_2)\sim~$5.5~K, in excellent agreement with the measured value. The single-ion anisotropy parameters $C_2S^2=-0.4$~K and $C_4S^4=0.455$~K were chosen to reproduce our experimental phase topology, especially the transition at $ T_{ab}~\sim~$3~K.  The  phase diagrams obtained from a two-sublattice mean-field theory with magnetic field applied along the [100] and [001] directions are shown in Figs.~\ref{Fig4}b and c.  Although our calculated $c$-AF region persists to slightly higher fields than our data, the overall agreement is reasonably good considering the limitations of a mean field approach.

To the best of our knowledge, {\E} is the first example of a material undergoing a transition from a crystalline to disordered state at low temperature, despite lacking any intrinsic structural, magnetic or electronic frustration.  It is natural to question the effect of this disorder on the dielectric properties of {\E}: although neither the structural transition nor the AF order alone provide the spatial inversion symmetry breakage necessary to stabilise ferroelectricity, it is plausible that local disorder could cant the Eu$^{2+}$ moments via the DM interaction.  This would drive a local symmetry-breakage and hence enable the formation of nanoscale electric dipoles.  Structural disorder implies that the global dipole moment would average to zero, rendering FE order difficult to detect.  However, the possibility of a static polarization extending over useful lengthscales in less-disordered {\E} merits further attention.  

We thank I. Martin and R. Lortz for useful discussions.  This work was supported by The National Research Foundation of Singapore, the European Union through MEXT-CT-2006-039047 and EURYI research grants.  Work at the LANL was performed under the auspices of the U.S. DOE contract No. DE-AC52- 06NA25396 through the LDRD program.

\end{document}